\documentclass[showpacs,preprintnumbers,aps,prd,nofootinbib]{revtex4} 
\usepackage[dvips]{graphicx} 
\usepackage{bm,latexsym,amsmath,amssymb,amsfonts}

\begin{document} 

\title{Quantization of scalar perturbations in brane-world inflation} 

\author{Hiroyuki Yoshiguchi$^1$} 
\author{Kazuya Koyama$^{1,2}$} 

\address{$^1$ Department of Physics, University of Tokyo 7-3-1 Hongo, 
Bunkyo, Tokyo 113-0033, Japan} 
\address{$^2$ Institute of Cosmology and Gravitation, Portsmouth 
University, Portsmouth, PO1 2EG, UK}

\begin{abstract} 
We consider a quantization of scalar perturbations about a de Sitter brane 
in a 5-dimensional anti-de Sitter (AdS) bulk spacetime. 
We first derive the second order action for a master variable 
$\Omega$ for 5-dimensional gravitational perturbations. 
For a vacuum brane, there is a continuum of 
normalizable Kaluza-Klein (KK) modes with $m>3H/2$. 
There is also a light radion mode with $m=\sqrt{2}H$ which satisfies the 
junction conditions for two branes, but is non-normalizable for a 
single brane model. We perform the quantization of these bulk 
perturbations and calculate the effective energy density of the 
projected Weyl tensor on the barne. 
If there is a test scalar field perturbation on the brane, 
the $m^2 = 2H^2$ mode together with the zero-mode and an infinite ladder 
of discrete tachyonic modes become normalizable in a single brane model. 
This infinite ladder of discrete modes as well as the continuum of 
KK modes with $m>3H/2$ introduce corrections to the scalar field 
perturbations at first-order in a slow-roll expansion. 
We derive the second order action for the Mukhanov-Sasaki variable coupled to 
the bulk perturbations which is needed to perform the quantization and 
determine the amplitude of scalar perturbations generated 
during inflation on the brane. 
\end{abstract} 



\maketitle 

\section{Introduction} 
In recent years, there has been a great deal of interests in the 
brane-world picture where ordinary matter fields are confined to a 
lower-dimensional hypersurface (brane) embedded in a higher-dimensional 
bulk spacetime (see \cite{review,Reviews1,Reviews2,Reviews3} for reviews). 
In particular, models proposed by Randall and Sundrum are very 
interesting \cite{RSI,RSII}. 
In their second model (RS model), they showed that the conventional 
4-dimensional gravity can be recovered at low energies on a Minkowski 
brane with positive tension in 5-dimensional anti-de Sitter (AdS) 
spacetime \cite{RSII}. 
An interesting point is that the 4-dimensional gravity can be recovered 
despite the infinite size of the extra-dimension. 

At low energies, effects of the extra-dimension must be small from the 
observational point of view, but at high energies the effects could be 
dominant. 
The effects of the extra-dimension could affect generations of primordial 
fluctuations in the period of inflation in the early universe. 
Such imprints left on the spectrum of primordial fluctuations will be 
constrained by high precision observations of the Cosmic Microwave 
Background (CMB). 

In general, it is difficult to find analytic solutions for 
cosmological perturbations that properly satisfy the junction conditions 
on the brane \cite{KoyamaSoda2,koyama04a}. 
However, this can be done in the special case of a de Sitter brane in 
a 5-dimensional AdS bulk spacetime \cite{kaloper99,garriga00}. 
This fact is useful since it provides a zero-th order approximation for 
slow-roll inflation on the brane. 
The behavior of vector \cite{bridgman01} and tensor 
\cite{langlois00,Rubakov,fk,Tanaka} 
perturbations have been discussed already .

The amplitude of scalar perturbations excited by inflaton 
fluctuations on the brane is also computed in the extreme slow-roll 
limit where the coupling between inflaton field fluctuations and bulk 
perturbations can be neglected \cite{maartens00,Liddle}. 
However, to go beyond the zero-th order slow-roll approximation, we have 
to solve the bulk metric perturbations. 
This issue is recently discussed by Koyama et al.\cite{koyama04} within 
a classical theory, using a master variable $\Omega$ for gravitational 
perturbations in a 5-dimensional AdS spacetime\cite{Mukohyama,Kodama}. 
They showed that the bulk perturbations introduce corrections to the scalar 
field fluctuations at first order computed in a slow-roll expansion. 
Thus, to quantitatively calculate the amplitude of scalar perturbations 
at this order, we need to discuss a quantum theory of the bulk 
gravitational field as well as the inflaton field on the brane. 

In this paper, we study the quantization of bulk scalar metric 
perturbations about a de Sitter brane in the RS model. 
We review the background spacetime in Sec.\ref{review}. 
In Sec.\ref{master_all}, we then introduce the master variable $\Omega$ 
for gravitational perturbations in 5-dimensional AdS spacetime 
\cite{Mukohyama,Kodama} and derive its second order action 
(\ref{action_Omega}). Then 
we first consider quantum scalar perturbations in the absence of matter 
fields on the brane in Sec.\ref{quantum_graviton}. 
In this case, there is a continuum of normalizable KK modes with 
$m>3H/2$. 
These bulk perturbations are felt on the brane through the projection of 
the perturbed 5-dimensional Weyl tensor $\delta E_{\mu\nu}$. 
We calculate the vacuum expectation value of its effective energy 
density by using the second order action for $\Omega$. 
A light radion mode with $m=\sqrt{2}H$ also satisfies the junction 
condition on the branes, and is normalizable for a two branes model 
\cite{koyama04,GenSasaki,GenSasaki2}. 
The effective energy density of $\delta E_{\mu\nu}$ due to this 
radion mode is also calculated. 
Next we consider bulk scalar perturbations excited by scalar field 
fluctuations on the brane in Sec.\ref{scalar}. 
We first discuss the effect of the metric backreaction on the 
scalar field fluctuations and review the solution for the bulk 
perturbations derived in Ref. \cite{koyama04} in 
Sec. \ref{scalar_classical}. 
We show that the evolution equation for the Mukhanov-Sasaki variable has 
a correction term that comes from the bulk perturbations. 
To quantize the Mukhanov-Sasaki variable, we need its action coupled to 
the bulk gravitational field. 
We derive this action (\ref{Q_action}) in Sec.\ref{scalar_quantum}. 
We summarize the main results and discuss them in Sec. \ref{summary}.

\section{COSMOLOGICAL BACKGROUND} 
\label{review} 

The action describing the RS brane world is given by 
%
%
\begin{eqnarray} 
S = \int d^5 x \sqrt{-{}^{(5)}g} \left[ {1 \over 2 
\kappa^2} \left( {}^{(5)}R  +12\mu^2 \right) \right] 
 + \int d^4 x \sqrt{-g} 
\left[ -\lambda  +{\cal L}_{\rm matter} 
+\frac{1}{\kappa^2} K \right] , 
\label{action} 
\end{eqnarray} 
%
where $\kappa^2$ is 5-dimensional gravitational constant, 
and $\mu$ is the curvature scale of the AdS spacetime.  
The brane has tension $\lambda$ which is set to be $6\mu/\kappa^2$ in 
the RS model. 
The induced metric on the brane is denoted as $g$, and given by 
%
$ 
g_{AB} = {}^{(5)}g_{AB} - n_{A}n_{B}, 
$ 
%
%
where $n^A$ is the unit vector normal to the brane. 
Matter field is confined on the 3-brane and is described by the 
Lagrangian ${\cal L}_{\rm matter}$. 
The final term in the action is the Gibbons-Hawking term and 
$K$ is the trace of the extrinsic curvature of the brane. 
We assume the $Z_2$ symmetry across the brane.

The 5-dimensional Einstein equations are obtained by minimizing the 
action with respect to variations of the bulk metric: 
%
%
\begin{equation} 
 \label{5DEinstein} 
{}^{(5)}G_{AB} + {}^{(5)}g_{AB}\, \Lambda_5 = 0 \,. 
\end{equation} 
%
%
The 4-dimensional matter fields determine the brane 
trajectory in the AdS bulk through the junction condition by 
producing the jump in the extrinsic curvature at the brane. 
The surface energy-momentum on the brane can be 
split into two parts, $T_{\mu\nu}-\lambda g_{\mu\nu}$, where 
$T_{\mu\nu}$ is taken to be the matter energy-momentum tensor and 
$\lambda$ a constant brane tension. 
The junction condition is then~\cite{BDL,SMS} 
\begin{equation} 
\label{Israel} 
 \left[K^\mu_\nu\right] \equiv K^{\mu~+}_\nu - 
K^{\mu~-}_\nu = 
 -\kappa_5^{~2} \left( T^\mu_\nu-{1\over 3} g^\mu_\nu \left( T - 
     \lambda \right) \right), 
\end{equation} 
where the extrinsic curvature of the brane is denoted by 
%
$ 
K_{\mu\nu}= g_\mu^{~C} g_\nu^{~D} \left({}^{(5)}\nabla_C n_D\right) \,, 
$ 
%
where ${}^{(5)}\nabla_C$ is the 5D covariant derivative. 
The induced Einstein equations on the brane are then given by 
\cite{SMS} 
%
%
\begin{eqnarray} 
\label{4DEinstein} 
G_{\mu\nu} = \kappa_4^2 T_{\mu\nu} + \kappa^4 \Pi_{\mu\nu} -E_{\mu\nu} 
\,, 
\end{eqnarray} 
%
%
where $\kappa_4^2 = \mu \kappa^2$. 
This equation include the projected five-dimensional Wey tensor 
$E_{\mu\nu}$ as well as the high energy correction term $\Pi_{\mu\nu}$ 
which is quadratic in the energy-momentum tensor $T_{\mu\nu}$. 

In this paper, we consider inhomogeneous bulk metric perturbations in 
the special case of a de Sitter brane in the AdS bulk. 
This is a good approximation to a brane-world inflation model with a 
slow-rolling scalar field on the brane. 
The metric of the unperturbed 5D spacetime in terms of the conformal 
bulk-coordinate is given by 
%
\begin{eqnarray} 
ds^2 &=& e^{2 W(z)} \biggl[ dz^2  
-dt^2 + e^{2 \alpha(t)}\delta_{ij} dx^i dx^j 
\biggr], 
\end{eqnarray} 
%
where 
%
\begin{eqnarray} 
e^{\alpha(t)} = {\rm exp} (Ht), \,\, e^{W(z)} = \frac{H}{\mu {\rm sinh}Hz}, 
\end{eqnarray} 
%
and $H(=\dot\alpha(t))$ is a constant Hubble expansion rate on the brane. 
The Cauchy horizon is located at $|z|=\infty$, and the brane is located at 
$z=z_0$ given by
%
\begin{eqnarray} 
z_0 = \frac{1}{H} \sinh^{-1} \frac{H}{\mu}. 
\end{eqnarray} 
%
The background metric satisfies 
%
\begin{eqnarray} 
&&H^2 + W'' -W'^2 =0, 
\\ 
&&W'' = \mu^2 e^{2W}. 
\end{eqnarray} 
%

%
%

\section{MASTER VARIABLE AND SECOND ORDER ACTION} 
\label{master_all}

\subsection{Master variable} 
\label{master} 
Here we introduce the master variable for gravitational perturbations in 
the 5-dimensional AdS spacetime firstly found by Mukohyama 
\cite{Mukohyama,Kodama}. 
The perturbed metric is given by~\cite{BMW} 
%
\begin{eqnarray} 
ds^2 &=& e^{2 W(z)} \biggl[ (1+2A_{yy})dz^2  
+2 A_y   dt dz  
-(1+2 A )dt^2  
\nonumber\\&& 
+ e^{2 \alpha(t)} \biggl((1+2{\cal R} )\delta_{ij} dx^i dx^j  
+ 2E_{,ij} dx^i dx^j + 2B_{,i} dx^i dt + 2C_{,i} dx^i dz \biggr) 
\biggr]. 
\end{eqnarray} 
%
Under a scalar type gauge transformation, 
%
\begin{eqnarray} 
&&t \to \bar{t} = t + \xi^t Y, 
\\ 
&&z \to \bar{z} = z + \xi^z Y, 
\\ 
&&x^i \to \bar{x}^i = x^i + \xi^S Y_,^i, 
\end{eqnarray} 
%
the metric variables transform as 
%
\begin{eqnarray} 
A_{yy} &\to& \bar{A}_{yy} = A_{yy}-\xi^z {}' -W'\xi^z, 
\nonumber\\ 
A_y &\to& \bar{A_y} = A_y + \xi^t {}'-\dot{\xi}^z, 
\nonumber\\ 
C &\to& \bar{C}=C - e^{-2\alpha}\xi^z - \xi^S{}', 
\nonumber\\ 
A &\to& \bar{A}=A-W'\xi^z -\dot{\xi}^t, 
\nonumber\\ 
B &\to& \bar{B}=B + e^{-2\alpha}\xi^t - \dot{\xi}^S, 
\nonumber\\ 
E &\to& \bar{E}=E - \xi^S, 
\nonumber\\ 
{\cal R} &\to& \bar{\cal R}={\cal R} -W'\xi^z -\dot{\alpha}\xi^t, 
\label{gauge_ads} 
\end{eqnarray} 
%
where dot and prime denote the derivative with respect to $t$ and $z$, 
respectively. 
In order to eliminate the gauge dependence on choice of 3-space 
coordinates we introduce the spatially gauge-invariant combinations 
%
%
\begin{equation}  
\begin{array}{lcl} 
 \sigma_t &\equiv& -B + \dot{E} \, ,\\ 
 \sigma_z &\equiv& -C + E' \, , 
\end{array} 
\label{scalarGI} 
\end{equation} 
%
%
which are subject only to temporal and bulk gauge transformations. 

The bulk and temporal gauges are fully determined by setting 
$\sigma_t=\sigma_z=0$, which has been termed the 5D-longitudinal 
gauge~\cite{cvdb,BMW}. 
We can define the remaining metric perturbations in the 
5D-longitudinal gauge as 
%
%
\begin{eqnarray} 
\Phi &=& A - \left(e^{2\alpha}\sigma_t \right)^\cdot 
 +W' e^{2\alpha} \sigma_z \, , \nonumber \\ 
\Psi &=& {\cal R} - \dot\alpha e^{2\alpha} \sigma_t 
+ W' e^{2\alpha} \sigma_z \, , \nonumber \\ 
S &=& A_y + e^{2\alpha}\sigma_t' 
+ \left( e^{2\alpha} \sigma_z\right)^\cdot \, , 
\nonumber \\ 
N &=& A_{yy} +e^{2\alpha} \left( \sigma_z' +W'\sigma_z\right) \, . 
\end{eqnarray} 
%
These are equivalent to the gauge-invariant bulk perturbations 
originally introduced in covariant form by 
Mukohyama~\cite{Mukohyama,Kodama} and in a coordinate-based approach by 
van den Bruck et al~\cite{cvdb}. 
These metric variables satisfy the three constraint equations, 
%
\begin{eqnarray} 
&&N + \Phi + \Psi = 0, 
\label{const1} 
\\ 
&&-\Phi'-2{\Psi}' +3W'N -\frac{1}{2}\left({\dot S} 
+H S \right)  = 0, 
\label{const2} 
\\ 
&&-{\dot N} + HN -2{\dot \Psi} +2H\Phi 
+\frac{1}{2}\left(S' +3W'S\right) 
 = 0. 
\label{const3} 
\end{eqnarray} 
%

In \cite{Mukohyama} (see also Ref.~\cite{Kodama}), it was 
shown that the perturbed 5D Einstein equations, 
in the absence of bulk matter perturbations are solved in an AdS 
background using a ``master variable'', $\Omega$. 
In the special case of a de Sitter brane in the AdS bulk, the 
metric variables are written by the master variable $\Omega$ as 
%
\begin{eqnarray} 
{\Phi} &=& -\frac{e^{-\alpha-3W}}{6} \biggl( 2\Omega'' 
-3W'\Omega' +{\ddot\Omega} -\mu^2 e^{2W}\Omega \biggr), 
\label{metric_omega_phi} 
\\ 
{S} &=& e^{-\alpha-3W} \biggl( {\dot\Omega}' -W'{\dot\Omega} 
\biggr), 
\label{metric_omega_s} 
\\ 
{N} &=& \frac{e^{-\alpha-3W}}{6} \biggl( \Omega'' 
-3W'\Omega' +2{\ddot\Omega} +\mu^2 e^{2W}\Omega \biggr), 
\label{metric_omega_n} 
\\ 
{\Psi} &=& \frac{e^{-\alpha-3W}}{6} \biggl( \Omega'' 
-{\ddot\Omega} -2\mu^2 e^{2W}\Omega \biggr). 
\label{metric_omega_psi} 
\end{eqnarray} 
%
These expressions satisfy the constraint 
equations Eq.(\ref{const1}-\ref{const3}).

Now, we derive the equation of motion for the master 
variable $\Omega$. 
By substituting the expression 
Eq.(\ref{metric_omega_phi}-\ref{metric_omega_psi}) 
into the dynamical parts of the perturbed Einstein equations, 
we can show that they are equivalent to \cite{Mukohyama} 
%
\begin{eqnarray} 
\Delta_{(S)}'' -W'\Delta_{(S)}' -\mu^2 e^{2W}\Delta_{(S)}=0, 
\nonumber\\ 
\ddot\Delta_{(S)}-W'\Delta_{(S)}' +\mu^2 e^{2W}\Delta_{(S)}=0, 
\nonumber\\ 
\dot\Delta_{(S)}' -W'\dot\Delta_{(S)} =0, 
\label{delta_eq} 
\end{eqnarray} 
%
where 
%
\begin{eqnarray} 
\Delta_{(S)} = 
e^{2\alpha} \left[{\ddot\Omega} -3H \dot\Omega -\left(\Omega'' 
-3W'\Omega'\right) -\mu^2 e^{2W}\Omega -e^{-2\alpha}\Delta \Omega 
\right]. 
\label{triangle_def} 
\end{eqnarray} 
%
Here $\Delta$ is the 3D spatial Laplacian 
$\Delta\Omega = \delta^{ij}\Omega_{,ij}$. 
Then, we can obtain the following equation from 
Eq.(\ref{delta_eq}):
%
\begin{eqnarray} 
\ddot{\tilde\Omega} -3H \dot{\tilde\Omega} -\left(\tilde\Omega'' 
-3W'\tilde\Omega'\right) -e^{-2\alpha}\Delta\tilde\Omega 
-\mu^2 e^{2W}\tilde\Omega =0, 
\label{eq_omega2} 
\end{eqnarray} 
%
where  
%
\begin{eqnarray} 
\Delta \tilde\Omega &=& \Delta \Omega 
+\Delta_{(S)}. 
\label{tilde_omega_def} 
\end{eqnarray} 
%
Here we note that there is a symmetry between $\Omega$ and 
$\tilde\Omega$. It is possible to show that a replacement of $\Omega$ with $\tilde\Omega$ 
in Eq.(\ref{metric_omega_phi}-\ref{metric_omega_psi}) 
does not alter the metric variables if the perturbed 
Einstein equations Eq.(\ref{delta_eq}) are satisfied. 
Then we can set $\Delta_{(S)}=0$ if $\Delta \Omega \neq 0$
as is shown in Ref. \cite{Mukohyama}. This is equivalent to say that the solutions 
of the metric variables are given by Eq.(\ref{metric_omega_phi}-\ref{metric_omega_psi}) 
where 
$\Omega$ is a solution of the master equation: 
%
\begin{eqnarray} 
\ddot{\Omega} -3H \dot{\Omega} -\left(\Omega'' 
-3W'\Omega'\right) -e^{-2\alpha}\Delta\Omega 
-\mu^2 e^{2W}\Omega =0. 
\label{eq_omega} 
\end{eqnarray} 
%
We also see that $\tilde\Omega$ coincides with $\Omega$ if we use 
Eq.(\ref{eq_omega}).

\subsection{Second order action for master variable} 
\label{omega}

In this subsection, we derive the second order action for $\Omega$ 
including the surface terms on the brane. 
By perturbing the gravitational part (except for ${\cal L}_{\rm matter}$) 
of the action Eq.(\ref{action}) up to second order, we get 
%
\begin{eqnarray} 
\delta_2 S  
&=& \int d^5 x 
\frac{e^{3\alpha + 3W}}{\kappa^2} \biggl[  
3(H^2 + W'' + 3W'{}^2 )A_{yy}^2 
-6H^2 A^2 -6H^2 A_{yy} A -3H^2 A_y^2 
\nonumber\\&& 
-2e^{-2\alpha} (A\Delta A_{yy} + 2{\cal R}\Delta A_{yy} + 
{\cal R}\Delta{\cal R} + 2A\Delta{\cal R} ) 
\nonumber\\&& 
-6 W'A' A_{yy}  +6H \dot A_{yy}A +6 W'\dot A_{yy} A_y  
+6H A' A_y -\frac{1}{2}e^{-2\alpha} A_y\Delta A_y 
\nonumber\\&& 
-2A_y (3\dot{\cal R}' +\Delta \dot{E}' -\Delta \dot{C}) 
-\frac{1}{2} e^{2\alpha} (B'-\dot{C} -2e^{-2\alpha}A_y) \Delta 
(B'-\dot{C}) 
\nonumber\\&& 
-6{\cal R}'{}^2 +6\dot{\cal R}^2 
-2(3\dot{\cal R} +\Delta\dot{E} -\Delta B) (\dot{A}_{yy} -H A_{yy} 
+2\dot{\cal R} -2H A) 
\nonumber\\&& 
+2(3{\cal R}' +\Delta E' -\Delta C) (A' +2{\cal R}' 
-3W'A_{yy} -H A_y ) 
\biggr]. 
\label{action_metric} 
\end{eqnarray} 
%
Here any gauge conditions are not imposed. 
The surface term on the brane cancels out. 
This action can be simplified by imposing the 5D-longitudinal gauge and 
using the constraint equations Eq.(\ref{const1})-(\ref{const3}) as 
%
\begin{eqnarray} 
\delta_2 S  
&=& \int d^5 x 
\frac{e^{3\alpha + 3W}}{\kappa^2} \biggl[  
3(H^2 + W'' + 3W'{}^2 ) N^2 
-6H^2 \Phi^2 -6H^2 N \Phi -3H^2 S^2 
\nonumber\\&& 
-2e^{-2\alpha} (\Phi\Delta N + 2\Psi\Delta N + 
\Psi\Delta\Psi + 2\Phi\Delta\Psi ) 
\nonumber\\&& 
-6 W'\Phi' N  +6H \dot N \Phi +6 W'\dot N S  
+6H \Phi' S -\frac{1}{2}e^{-2\alpha} S\Delta S 
-6\Psi'{}^2 +6\dot\Psi^2 
\biggr] 
\nonumber\\&& 
+\int d^4 x \frac{e^{3\alpha}}{\kappa^2} \left[ 
3S \dot\Psi \right]. 
\label{action_metric_simple} 
\end{eqnarray} 
%

To obtain the second order action for $\Omega$, we first substitute the 
expressions of the metric variables in terms of $\Omega$, 
Eq.(\ref{metric_omega_phi})-(\ref{metric_omega_psi}). 
After integrating by parts, we get the following action 
%
\begin{eqnarray} 
S &=& \int d^5 x \frac{e^{\alpha -3W}}{6\kappa^2} \biggl[ 
\dddot\Omega^2 - \Omega'''^2 +3\dot\Omega''^2 -3\ddot\Omega'^2 
+e^{-2\alpha} \biggl( \ddot\Omega\Delta\ddot\Omega 
+ \Omega'' \Delta\Omega'' -2 \dot\Omega'\Delta\dot\Omega' 
\biggr) 
\nonumber\\&&~~~ 
+(14H^2 +3W'') \ddot\Omega^2 -2(H^2 +3W'') \Omega''^2 
+3(-4H^2 + W'') \dot\Omega'^2 
\nonumber\\&&~~~ 
+ 3(7W''^2 -14W'^2W'' +8W'^4)\dot\Omega^2 
+ (5W''^2 -22W'^2W'' +8W'^4)\Omega'^2 
+ 9W''^3\Omega^2 
\nonumber\\&&~~~ 
+2(3H^2 + W'') e^{-2\alpha} \dot\Omega\Delta\dot\Omega 
+\left(-2H^2 +W'' \right)e^{-2\alpha} \Omega' \Delta\Omega' 
+W''^2 e^{-2\alpha}\Omega\Delta\Omega 
\biggr] 
%
%
\nonumber\\&& 
- \int d^4 x \frac{e^{\alpha}}{6\kappa^2} \biggl[ 
5 \ddot\Omega' \Omega'' -\ddot\Omega'\ddot\Omega  
+4W' \Omega''^2 +4W' \ddot\Omega^2 -5W' \Omega''\ddot\Omega 
+\frac{3}{2}W' \dot\Omega'^2 -H \dot\Omega' \ddot\Omega 
-6H\dot\Omega' \Omega'' 
\nonumber\\&&~~~ 
+(4W'' -6W'^2) \Omega'\Omega'' +6HW' \dot\Omega \Omega'' 
-(11H^2 +5W'') \dot\Omega'\dot\Omega -2W'W'' \Omega\Omega'' 
\nonumber\\&&~~~ 
+W' \left(-\frac{3}{2}H^2 + 4W'' \right) \Omega'^2 
+W' \left( 12H^2 -\frac{11}{2} W''\right) \dot\Omega^2 
-H \left( 17H^2 +22W'' \right) \dot\Omega\Omega' 
\nonumber\\&&~~~ 
+W'' \left( 9H^2 +8W'' \right) \Omega\Omega' 
+W'W'' \left( \frac{5}{2}H^2 -5W'' \right) 
\nonumber\\&&~~~ 
+e^{-2\alpha} \left( 
-\dot\Omega' \Delta \dot\Omega +H\Omega'\Delta\dot\Omega 
-W'' \Omega'\Delta\Omega -\frac{1}{2}W'W'' \Omega\Delta\Omega 
-\frac{3}{2}W' \Omega'\Delta\Omega' +3W'\dot\Omega\Delta\dot\Omega 
\right) 
\biggr]. 
\end{eqnarray} 
%
Although this action is extremely long, we can summarize the result in 
the following form: 
%
\begin{eqnarray} 
&&S = \int d^5 x \frac{e^{-3\alpha - 3W}}{6\kappa^2} \biggl[ 
\left( \Delta\dot{\tilde\Omega} \right)^2  
- \left( \Delta\tilde\Omega' \right)^2 
+e^{-2\alpha}\Delta\tilde\Omega\Delta^2\tilde\Omega  
+\mu^2 e^{2W}\left( 
\Delta\tilde\Omega \right)^2 \biggr] 
\nonumber\\&&~~~ 
+\int d^4 x \frac{e^{\alpha}}{6\kappa^2} \biggl[ 
\frac{9}{2}W' \dot {\cal F}^2 
-\frac{3}{2}W' e^{-2\alpha} {\cal F}\Delta {\cal F} 
- W'e^{-4\alpha} \left(\Delta \tilde\Omega \right)^2 
-3 \ddot{\cal F} e^{-2\alpha}\Delta (\Omega -\tilde\Omega) 
\biggr], 
\label{action_Omega2} 
\end{eqnarray} 
%
where 
%
\begin{eqnarray} 
{\cal F} &=& \Omega' -W'\Omega, 
\end{eqnarray} 
%
and $\tilde{\Omega}$ is defined in Eq. (\ref{tilde_omega_def}).
This action contains higher derivative terms with respect to $t$.  
In order to perform a quantization 
of $\Omega$, we need an action that contains up to second derivatives. 
This can be achieved using the symmetry between
$\tilde{\Omega}$ and $\Omega$. It is possible to replace
$\tilde\Omega$ with $\Omega$ or equivalently we can set 
$\Delta_{(S)}=0$ without losing the physical degrees of 
freedom \cite{Mukohyama}. Then we can derive the second order action for $\Omega$
that contains up to second derivatives with respect to $t$: 
%
\begin{eqnarray} 
&&S = \int d^5 x \frac{e^{-3\alpha - 3W}}{6\kappa^2} \biggl[ 
\left( \Delta\dot{\Omega} \right)^2  
- \left( \Delta\Omega' \right)^2 
+e^{-2\alpha}\Delta\Omega\Delta^2\Omega  
+\mu^2 e^{2W}\left( 
\Delta\Omega \right)^2 \biggr] 
\nonumber\\&&~~~ 
+\int d^4 x \frac{e^{\alpha}}{6\kappa^2} \biggl[ 
\frac{9}{2}W' \dot {\cal F}^2 
-\frac{3}{2}W' e^{-2\alpha} {\cal F}\Delta {\cal F} 
- W'e^{-4\alpha} \left(\Delta\Omega \right)^2 
\biggr]. 
\label{action_Omega} 
\end{eqnarray} 
%
Defining 
%
\begin{eqnarray} 
\omega = \frac{e^{-3\alpha-3W}}{\sqrt{3\kappa^2}} \Delta \Omega, 
\end{eqnarray} 
%
the second order action for $\Omega$ in the bulk can be rewritten into 
the form of a 5-dimensional scalar field: 
%
\begin{eqnarray} 
S &=& \frac{1}{2} \int d^5 x e^{3\alpha +3W} \biggl[ 
\dot\omega^2 -\omega'^2 + e^{-2\alpha} \omega\Delta\omega  
+ 4W'' \omega^2 \biggr]. 
\end{eqnarray} 
%
This form of the action is convenient when we fix the normalization of 
$\Omega$. 

\subsection{General solution for master variable} 
In this subsection, we derive the general solution for $\Omega$. 
Solutions of the wave equation Eq.(\ref{eq_omega}) can be separated into 
eigenmodes of the time-dependent equation on the brane and bulk mode 
equation: 
%
\begin{eqnarray} 
\Omega(t,y;\vec{x})= \int d^3\vec{k}\, dm\, v_m(t) u_m(y) 
e^{i\vec{k}.\vec{x}} \,, 
\end{eqnarray} 
%
where 
%
\begin{eqnarray} 
\ddot v_m -3H\dot v_m+\left[ m^2+{k^2 e^{-2\alpha}}\right] 
 v_m &=&0\,, \label{varphieom}\\ 
u_m''-3 W' u_m'+ \left[ m^2 +  \mu^2 e^{2W}\right] u_m &=& 0\,. 
\label{bulkmodeeq} 
\end{eqnarray} 
%
The general solution for $v_m$ is given by 
\begin{equation} 
\label{varphisol} 
v_m(\eta;\vec{k}) = \left(-k\eta\right)^{-3/2}\, B_\nu(-k\eta)\,, ~~ 
\nu^2={9\over4}-{m^2\over H^2}\,, 
\end{equation} 
where $B_\nu$ is a linear combination of Bessel functions of order 
$\nu$. 

Defining $\Psi_m\equiv e^{-3W/2} u_m$, it is 
possible to rewrite the off-brane equation (\ref{bulkmodeeq}) in 
Schr\"odinger-like form: 
%
\begin{equation} 
\label{SE} 
 {d^2\Psi_m\over dz^2} - V\Psi_m =-m^2 \Psi_m \,, 
\end{equation} 
%
where 
%
\begin{eqnarray} 
V(z)&=&- \frac{1}{4}\mu^2 e^{2W(z)}+{{9\over4}}H^2 
\,,\nonumber\\ &=& - {H^2 \over 4\sinh^2(Hz)} + {{9\over4}}H^2 
 \,. 
\end{eqnarray} 
%
For $z\to\infty$ we have $V\to9H^2/4$ and we have a continuum of 
massive modes for $m^2>9H^2/4$ which become oscillating plane waves 
as $z\to\infty$. 
The general solution of the mode function in the $z$-direction is 
%
\begin{eqnarray} 
u_m (z) = \left( \sinh Hz\right)^{-1} W_{\nu-1/2} (\cosh Hz), 
\end{eqnarray} 
%
where $W_\alpha$ is a linear combination of Legendre functions of order 
$\alpha$. 


\section{QUANTUM GRAVITON WITH A VACUUM DE SITTER BRANE} 
\label{quantum_graviton} 
We first consider a quantization of scalar perturbations 
in the absence of matter perturbations.  
If we take the variation of the action Eq.(\ref{action_Omega}) with 
respect to $\Omega$, we get the equation of motion for $\Omega$ in the 
bulk Eq.(\ref{eq_omega}) and the junction condition on the brane 
%
\begin{eqnarray} 
{\cal F}=0. 
\label{vac_junc} 
\end{eqnarray} 
%
This is consistent with the condition obtained by minimizing 
the action with respect to ${\cal F}$. 

The effect of bulk perturbations is felt through the 
projected Weyl tensor $E_{\mu \nu}$ in Eq.(\ref{4DEinstein}).
In the background spacetime $E_{\mu \nu}=0$ and 
the energy density of perturbed projected Weyl tensor $\delta E_{\mu\nu}$ 
is given in terms of the master variable by \cite{Deffayet} 
%
\begin{eqnarray} 
\kappa_4^2 \delta \rho_E = \frac{k^4e^{-5\alpha}}{3} \Omega. 
\label{E_rho} 
\end{eqnarray} 
%
In order to estimate the effect of Weyl tensor, we will compare the Weyl energy density 
perturbation with the background energy density given by 
\begin{equation}
\rho_\Lambda = \frac{3H^2}{\kappa_{4,{\rm eff}}^2}, \quad 
\kappa_{4,{\rm eff}}^2 = \kappa_4^2 
\left[1+\left(\frac{H}{\mu}\right)^2 \right]^{1/2}.
\end{equation}
We define the power spectrum $P_E(k)$ of $\delta \rho_{E}$ normalized 
by $\rho_{\Lambda}$:
\begin{equation}
\frac{<\delta \rho_{E}^2>}{\rho_{\Lambda}^2} 
= \int \frac{dk}{k} P_E(k).
\end{equation}

\subsection{Quantization of heavy modes} 
\label{quantum_heavy} 
Here we consider the quantization of the heavy modes with 
$m > \frac{3}{2}H$. 
The junction condition for $\Omega$ on the brane 
is given by Eq.(\ref{vac_junc}). 
From this condition, the solution of the bulk mode function is written 
as 
%
\begin{eqnarray} 
u_m(z) &=& C(m) ({\rm sinh} Hz)^{-1} \biggl(P_{i\gamma-1/2} 
({\rm cosh}Hz) +\beta(m)Q_{i\gamma-1/2} ({\rm cosh}Hz) \biggr), 
\end{eqnarray} 
%
where 
%
\begin{eqnarray} 
\gamma &=& \sqrt{\frac{m^2}{H^2} -\frac{9}{4}}, 
\\ 
\beta(m) &=& -\frac{P^1_{i\gamma-1/2} ({\rm cosh}Hz_0)}
{Q^1_{i\gamma-1/2} ({\rm cosh}Hz_0)}, 
\end{eqnarray} 
%
where $P^{\mu}_{\nu}$ and $Q^{\mu}_{\nu}$ are associated Legendre 
functions of the first kind and second kind respectively. 
We can determine the coefficient $C(m)$ by the normalization condition \cite{KKS} 
%
\begin{eqnarray} 
2 \int_{z_0}^\infty d (Hz) e^{-3W} u_m (z) u_{m'}^* (z) 
= \delta (\gamma' - \gamma), 
\end{eqnarray} 
%
as 
%
\begin{eqnarray} 
C(m) = \left( \frac{H}{\mu} \right)^{3/2}  
\frac{1}{\sqrt{\zeta(m) + \xi(m)}}, 
\end{eqnarray} 
%
where 
%
\begin{eqnarray} 
\zeta(m) = \left|\frac{\Gamma(i\gamma)}{\Gamma(i\gamma + 1/2)}\right|^2, 
\,\, 
\xi(m) = \left|\frac{\Gamma(-i\gamma)}{\Gamma(-i\gamma + 1/2)} + \pi\beta(m) 
\frac{\Gamma(i\gamma+1/2)}{\Gamma(i\gamma + 1)} \right|^2. 
\end{eqnarray} 
%
The normalization of the time mode function is determined so that the correct canonical 
quantization of 
$\omega$ is ensured:
%
\begin{eqnarray} 
v_m (\eta) =  \frac{\sqrt{3\kappa^2}}{-k^2} 
\frac{\sqrt\pi}{2} 
 (-H\eta)^{-3/2} e^{-\gamma\pi/2} 
H_{i\gamma}^{(1)} (-k\eta). 
\end{eqnarray} 
%

Next we calculate the vacuum expectation value for the energy density of 
the projected Weyl tensor $\delta E_{\mu\nu}$ generated by the KK modes 
with $m>3H/2$. 
The vacuum expectation value of $\delta \rho_E^2$ is given by
%
\begin{eqnarray} 
\left< \left(\kappa_4^2 \delta \rho_E(x)\right)^2 \right> = 
\frac{1}{(2 \pi)^3}
\int d^3 k \frac{k^8}{9} e^{-10\alpha}  
 \int_0^\infty d \gamma \left| u_m(z_0) \right|^2 \left| v_m \right|^2. 
\label{omega_vac} 
\end{eqnarray} 
%
However we have to notice that in large $m$ and $k$ limit 
$\left| v_m \right|^2$ behaves 
%
\begin{eqnarray} 
\left| v_m \right|^2 \sim \frac{3\kappa^2}{k^4} 
\frac{(-H\eta)^{-3}}{2 \sqrt{(m/H)^2 + (-k\eta)^2}} 
~~~ 
(m,k \to \infty). 
\end{eqnarray} 
%
This shows that $m$ integral in Eq.(\ref{omega_vac}) has a
logarithmic divergence. 
This ultraviolet divergence appears in the 5-dimensional field theory 
even in Minkowski spacetime. 
Thus we have to subtract this divergence \cite{KKS}.
Then the correct vacuum expectation value becomes 
%
\begin{equation} 
\left< \left(\kappa_4^2 \delta \rho_E(x)\right)^2 \right> 
= \frac{1}{(2 \pi)^3}
\int d^3 k \frac{k^8}{9} e^{-10 \alpha} \int_{0}^\infty d \gamma 
\left| u_m(z_0) \right|^2 \left( \left| v_m \right|^2 
- \frac{3\kappa^2}{k^4} 
\frac{(-H\eta)^{-3}}{2\sqrt{(m/H)^2 + (-k\eta)^2}} 
\right).
\label{omega_vac_correct} 
\end{equation} 
%
Then the power spectrum $P_E(k)$ is given by
%
\begin{eqnarray} 
P_{E}(k) &=& 
\frac{1}{108 \pi^2} \left( 
\frac{k e^{-\alpha}}{H} \right)^7 
(\kappa_4 H)^2  C_{KK}^2, \\
C_{KK}^2 &=& \frac{H}{\mu} 
\left(1+ \left(\frac{H}{\mu}\right)^2 \right)
\int_{0}^\infty d \gamma \left| u_m(z_0) \right|^2 
\left( \frac{\pi}{2}e^{-\gamma\pi} 
\left| H_{i\gamma}^{(1)} (-k\eta)\right|^2 
-\frac{1}{\sqrt{(m/H)^2 + (-k\eta)^2}} \right). 
\label{kk_amp} 
\end{eqnarray} 
%
The amplitude of $C_{KK}^2$ is enhanced for large $H/\mu$ as expected. 

%
%

\subsection{Quantization of radion mode} 
Next we discuss the quantization of the radion mode 
considered in \cite{koyama04}. 
The vacuum junction condition Eq.(\ref{vac_junc}) is trivially satisfied 
for any $z$ by the bulk mode solution 
%
\begin{eqnarray} 
u_{\rm r} \propto e^W, m^2=m_{\rm r}^2=2H^2. 
\label{radion_bulk} 
\end{eqnarray} 
%
This mode is non-normalizable in the single brane model. 
But in the two brane model, where the second brane is located at any 
fixed $z_1 > z_0$, this mode is normalizable, and automatically 
satisfies the boundary condition at the second brane. 
This mode is identified as the radion in \cite{koyama04}. 
The time dependence of the radion mode is 
%
\begin{eqnarray} 
v_{\rm r} = (-k\eta)^{-3/2} B_{1/2} (-k\eta). 
\end{eqnarray} 
%

Substituting the bulk mode function for the radion 
Eq.(\ref{radion_bulk}) into the second order action for $\Omega$ 
Eq.(\ref{action_Omega}), we obtain the action for the time dependence of 
the radion as 
%
\begin{eqnarray} 
S_{\rm r} = \frac{N_k}{6\kappa^2} \int d^4 x e^{-3\alpha} \left( 
\left( \Delta\dot v_{\rm r} \right)^2 -e^{-2\alpha} \Delta v_{\rm r} 
\Delta^2 v_{\rm r} -H^2 \left( \Delta v_{\rm r} \right)^2 \right), 
\end{eqnarray} 
%
where 
%
\begin{eqnarray} 
N_k = \int_{z_0}^{z_1} dz e^{-W} 
= \int_{z_0}^{z_1} dz \frac{\mu\sinh Hz}{H}, 
\end{eqnarray} 
%
and we used the vacuum boundary condition for $\Omega$, ${\cal F} = 0$. 
If we define a canonical 4-dimensional field 
\begin{equation}
\psi_k = {\sqrt \frac{N_k}{3\kappa^2}} e^{-3\alpha}\Delta v_{\rm r}, 
\end{equation}
this action becomes the form of a 4-dimensional scalar field with $m^2 = 
2H^2$: 
%
\begin{equation} 
S_{\rm r} = \frac{1}{2} \int d^4 x e^{3\alpha} \left( 
\dot \psi_k^2 -e^{-2\alpha} \psi_k \Delta \psi_k -H^2 \psi_k^2 \right). 
\end{equation} 
%
The solution of $\psi_k$ is easily obtained including its normalization 
as 
%
\begin{eqnarray} 
\psi_k = \frac{\sqrt\pi}{2} H^{-1/2} (-H\eta)^{3/2} 
 H_{1/2}^{(1)}(-k\eta). 
\label{psi_radion} 
\end{eqnarray} 
%
Using Eq.(\ref{psi_radion}), the Weyl energy density perturbations 
for the quantum radion becomes 
%
\begin{equation} 
\kappa_4^2 \delta \rho_E = -\frac{\sqrt\pi}{2} 
{\sqrt \frac{\kappa^2}{3N_k}} k^2 (-H\eta)^{7/2} H^{-1/2} 
H_{1/2}^{(1)} (-k\eta). 
\end{equation} 
%
This agrees with the result obtained by Gen and Sasaki \cite{GenSasaki2}. 

A growing mode solution represents the tachyonic instability 
of de Sitter two branes. As is shown in Ref. \cite{GenSasaki2}, 
the tachyonic instability is not strong enough to cause 
gravitational instabilities on the brane in the sense that 
$P_E(k)$ does not grow. As noted in \cite{koyama04}, a decaying 
mode also has an interesting physical meaning. The 
decaying mode corresponds to the dark radiation perturbation 
that is associated with a small black hole in the bulk
\cite{yoshiguchi04}. After inflation, the dark radiaiton
perturbation becomes a growing mode and this can affect CMB because 
the dark radiation perturbation
induces isocurvature perturbations \cite{LMSW,kcmb}. The power spectrum 
$P_E(k)$ coming form the decaying mode determines the initial condition 
for this isocurvature perturbations. Taking the long wavelength 
limit $k e^{-\alpha}/H \to 0$, $P_E$ is given by
\begin{eqnarray}
P_E(k) &=& \frac{1}{108 \pi^2} 
\left( \frac{k e^{-\alpha}}{H} \right)^8 (\kappa_4 H)^2
C_{dark}^2, \\ 
C_{dark}^2 &=& \left( \frac{H}{\mu} \right)^2 
\left(1+ \left(\frac{H}{\mu}\right)^2 \right)
\frac{1}{\cosh Hz_1 -\cosh H z_0}.
\end{eqnarray}
Although this could be large at high energies and/or 
when the second brane is close to the physical brane,
the effect is negligible on large scales because the 
spectrum is highly blue tilted. However, in order to address 
observational consequences, a detailed analysis of the evolution 
of perturbations after inflation is needed. 

\section{SCALAR FIELD ON THE BRANE} 
\label{scalar} 

In this section we include the scalar field perturbation 
$\delta \phi$ on the brane. 
We assume that the potential of the scalar field $\phi$ 
confined to the brane is very flat, so that the scalar field is 
slow-rolling. 
In such a situation, we can treat the corrections to the evolution 
of scalar field fluctuation due to the metric backreaction perturbatively. 
We first consider the problem of quantization of scalar perturbations 
using equations of motion based on the results of Ref. \cite{koyama04}.
We show that the evolution equation for the Mukhanov-Sasaki variable has 
a correction term that comes from the bulk perturbations. 
To quantize the Mukhanov-Sasaki variable, we need its action coupled to 
the bulk gravitational field, which is derived in 
Sec.\ref{scalar_quantum}.

\subsection{Equation of motion for scalar perturbations} 
\label{scalar_classical}


We expand the scalar field perturbations in terms of  
a slow-roll parameter; 
%
\begin{equation} 
\delta \phi= \delta \phi_0+\delta \phi_1 +... 
\end{equation} 
%
The 0-th order of the scalar field fluctuation obeys the following 
equation of motion, 
%
\begin{equation} 
\delta\ddot\phi_0 +3H\delta\dot\phi_0 +k^2 e^{-2\alpha}\delta\phi_0 = 0. 
\label{eom_phi0} 
\end{equation} 
%
The metric perturbations are generated by the 0-th order fluctuation of 
the scalar field through the induced Einstein equations on the brane, 
%
\begin{eqnarray} 
3 H \dot{\Psi}-3 H^2 \Phi + k^2 e^{-2\alpha} \Psi 
&=& \frac{\kappa_{4,{\rm eff}}^2}{2}  
(\dot{\phi} \dot{\delta \phi}_0 +V' \delta \phi_0) 
+\frac{\kappa_4^2}{2} \delta\rho_E, 
\label{induced_tt} 
\\ 
H\Phi -\dot\Psi &=& \frac{\kappa_{4,{\rm eff}}^2}{2} \dot\phi 
 \delta\phi_0 -\frac{\kappa_4^2}{2} \delta q_E, 
\label{induced_ti} 
\\ 
-\ddot{\Psi}-3H \dot{\Psi}+H \dot{\Phi}+3 H^2 \Phi 
-\frac{1}{3} k^2 e^{-2\alpha} (\Psi+\Phi) 
&=& \frac{\kappa_{4,{\rm eff}}^2}{2}  
(\dot{\phi} \dot{\delta \phi}_0 -V' \delta \phi_0) 
+\frac{\kappa_4^2}{6} \delta\rho_E, 
\label{induced_ii} 
\\ 
-e^{-2\alpha}(\Psi+\Phi) &=& \kappa_4^2 \delta\pi_E, 
\label{induced_ij} 
\end{eqnarray} 
%
where 
%
\begin{eqnarray} 
\kappa_4^2 \delta q_E &=& \frac{k^2 e^{-3\alpha}}{3} 
\left( \dot\Omega -H\Omega \right), 
\\ 
\kappa_4^2 \delta \pi_E &=& \frac{e^{-3\alpha}}{2} 
\left( \ddot\Omega -H\Omega + \frac{k^2 e^{-2\alpha}}{3}\Omega \right), 
\end{eqnarray} 
%
and $\delta \rho_E$ is given by Eq.(\ref{E_rho}) 
These metric fluctuations in turn affect the dynamics of the 1st-order 
scalar field perturbation as 
%
\begin{equation} 
\ddot{\delta \phi}_1 + 3H \dot{\delta \phi}_1  
+ k^2 e^{-2\alpha} \delta \phi_1  
=-V'' \delta \phi_0 - 3 \dot{\phi}  
\dot{\Psi} + \dot{\phi} \dot{\Phi}-2 V' \Phi. 
\end{equation} 
%

To calculate the quantum fluctuation of $\delta\phi_1$, we need to 
determine the metric perturbations, including the normalization.
In the standard 4-dimensional cosmology, this can be fixed from the 
normalization of $\delta \phi_0$ by using the $(t,i)$ component of the 
4-dimensional Einstein equations. 
However, in the brane-world, we can not determine $\Phi$ and $\Psi$ 
without the solution of $\Omega$ 
because there are the contributions from the projected Weyl tensor 
$\delta E_{\mu\nu}$. 

Yet we can determine the normalization of a part of $\Phi$ and $\Psi$ 
only from $\delta \phi_0$. 
This can be shown as follows. 
First, we rewrite the expressions of $\Phi$ and $\Psi$, 
Eq.(\ref{metric_omega_phi}) and (\ref{metric_omega_psi}), 
using $\Omega$ and ${\cal F}$ as 
%
\begin{eqnarray} 
\Psi &=& \frac{e^{-\alpha-3W}}{6} \biggl[ 
3W'{\cal F} -3H (\dot\Omega -H \Omega) 
-e^{-2\alpha} \Delta\Omega \biggr], 
\label{psi_fo} 
\\ 
\Phi &=& \frac{e^{-\alpha-3W}}{6} \biggl[ 
-3W'{\cal F} -3\ddot\Omega +6H \dot\Omega -3H^2 \Omega 
+2e^{-2\alpha} \Delta\Omega \biggr]. 
\label{phi_fo} 
\end{eqnarray} 
%
Here we used the equation of motion for $\Omega$ to reduce the number of 
$z$-derivatives. 
Substituting these expressions into the induced Einstein equations 
(\ref{induced_tt})-(\ref{induced_ij}), 
we obtain the equations written only by $\cal F$ and $\delta \phi_0$: 
%
\begin{eqnarray} 
-3 H \dot{\cal F} - k^2 e^{-2\alpha} {\cal F} 
 = \kappa^2 e^{\alpha} (\dot{\phi} \dot{\delta \phi_0} + V'(\phi) \delta 
 \phi_0) \,, 
\label{junc_k1b} 
\\ 
\dot{\cal F} = \kappa^2 e^{\alpha} \dot{\phi} \delta \phi_0 \,, 
\label{junc_k2b} 
\\ 
\ddot{\cal F} +2 H \dot{\cal F} 
 = \kappa^2 e^{\alpha} (\dot{\phi} \dot{\delta \phi_0} -V'(\phi) \delta 
 \phi_0)  \,. 
\label{junc_k3b} 
\end{eqnarray} 
%
These are the same as 4-dimensional Einstein equations 
if we define the quantities $\Phi_4$ and $\Psi_4$ by
\begin{equation}
{\cal F}=-2 \frac{e^\alpha}{W'} \Phi_4 = 2 \frac{e^\alpha}{W'} \Psi_4. 
\label{4dpart}
\end{equation}
We can interpret that ${\cal F}$ represents the 4-dimensional part of the 
metric variables and $\Omega$ contributes to them as a 
5-dimensional correction. 
Using Eq.(\ref{junc_k2b}), we can fix the normalization of $\cal F$ from 
the quantization of $\delta \phi_0$ in the same manner as the 
4-dimensional cosmology. 
However, this is not sufficient to determine the amplitude of 
metric perturbations $\Psi$ and $\Phi$. This is the limitation of the 
4-dimensional effective theory. We must solve the bulk equation for 
$\Omega$.  

The solution for $\Omega$ which satisfies the boundary condition on the 
brane is obtained in Ref. \cite{koyama04}. 
Combining the junction conditions, Eq.(\ref{junc_k1b})-(\ref{junc_k3b}), 
we get an evolution equation for ${\cal F}$; 
%
\begin{equation} 
\ddot{{\cal F}}-  \left(H + 2 \frac{\ddot{\phi}}{\dot{\phi}} \right) 
\dot{{\cal F}} + k^2 e^{-2\alpha} {\cal F}=0. 
\label{eqFb} 
\end{equation} 
%
This gives the boundary condition for $\Omega$. 
The scalar field fluctuation $\delta \phi_0$ is written by ${\cal F}(t)$ 
as  
%
\begin{equation} 
\kappa^2 \delta \phi_0 = e^{-\alpha} \frac{\dot{{\cal F}}}{\dot{\phi}}. 
\end{equation} 
%
The evolution equation for ${\cal F}$ Eq.(\ref{eqFb}) is consistent with 
the equation of motion for $\delta \phi_0$, Eq.(\ref{eom_phi0}). 
Assuming that $\phi$ is slow-rolling $|\ddot\phi / \dot\phi| \ll H$, 
the solution for $\cal F$ is 
%
\begin{eqnarray} 
{\cal F(\eta)} = C_1 \frac{{\rm cos}(-k\eta)}{-k\eta} + C_2 
 \frac{{\rm sin}(-k\eta)}{-k\eta}. 
\label{F_solution} 
\end{eqnarray} 
%
This gives the boundary condition for $\Omega$. 
The solution for $\Omega$ in the bulk subject to this condition is given 
by 
%
\begin{eqnarray} 
\Omega(z,\eta) &=& C_1 \sqrt{2\pi} 
\sum_{l=0}^{\infty} (-1)^{l}\left(2l+\frac{1}{2} \right) 
\frac{(\sinh Hz)^{-1} Q_{2l}(\cosh Hz)} 
{\mu Q^{1}_{2l}(\cosh Hz_0)} 
(-k \eta)^{-3/2} J_{2l+1/2}(-k \eta) \nonumber\\ 
&& + C_2 \sqrt{2\pi} 
\sum_{l=0}^{\infty} (-1)^{l}\left(2l+\frac{3}{2} \right) 
\frac{(\sinh Hz)^{-1} Q_{2l+1}(\cosh Hz)} 
{\mu Q^{1}_{2l+1}(\cosh Hz_0)} 
(-k \eta)^{-3/2} J_{2l+3/2}(-k \eta). 
\label{solution_omega} 
\end{eqnarray} 
%
Then it is possible to calculate the next order scalar field 
perturbations $\delta \phi_1$.

In order to evaluate the effect from metric perturbations, 
it is useful to use Mukhanov-Sasaki variable defined by 
%
\begin{equation} 
Q =\delta \phi-\frac{\dot{\phi}}{H} \Psi. 
\end{equation} 
%
In terms of slow-roll expansion, we have $Q_0 = \delta \phi_0$ and 
$Q_1 = \delta \phi_1 - \frac{\dot\phi}{H} \Psi$. 
Then using the induced Einstein equations, Eq.(\ref{induced_tt}), 
Eq.(\ref{induced_ii}), and Eq.(\ref{induced_ij}), we can derive the 
equation for $Q_1$; 
%
\begin{equation} 
\ddot Q_1 + 3 H \dot Q_1 + k^2 e^{-2\alpha} Q_1 
=-V'' Q_0 - 6 \dot{H} Q_0 + J, 
\label{Qevolution} 
\end{equation} 
%
where  
%
\begin{eqnarray} 
J &=& -\frac{\kappa_4^2 \dot{\phi}}{3 H}  
\left(k^2 \delta\pi_E + \delta\rho_E \right) \nonumber\\ 
&=&-\frac{\dot{\phi}}{H}\frac{k^2 e^{-3\alpha}}{6} 
\left(\ddot{\Omega}-H \dot{\Omega} + k^2 e^{-2\alpha}\Omega \right).  
\label{correction_J} 
\end{eqnarray} 
%
The equation is the same as the standard 4-dimensional cosmology except 
for the term $J$. 
$J$ describes the corrections that comes from the  
5-dimensional bulk perturbations. In order to address the 
quantization of $Q$, we need the second order action for 
$Q$. 

\subsection{Second order action for Mukhanov-Sasaki variable} 
\label{scalar_quantum} 
From now on, we derive the second order action for Mukhanov-Sasaki 
variable $Q$ coupled to the bulk metric perturbations 
described by the master variable $\Omega$. 
We first add the action of the 4-dimensional scalar field, 
%
\begin{eqnarray} 
S &=& \int d^4 x \sqrt{-g_4} \biggl( 
-\frac{1}{2} \partial_\mu \phi \partial^\mu \phi 
-V( \phi ) 
\biggr), 
\label{4dscalar_action} 
\end{eqnarray} 
%
to the gravitational part of the action.
Perturbing the action Eq.(\ref{4dscalar_action}) up to second order, 
we get 
%
\begin{eqnarray} 
S &=& \frac{1}{2} \int d^4 x e^{3\alpha} \biggl[ 
\delta \dot\phi^2 + e^{-2\alpha} \delta\phi \Delta \delta\phi 
-V'' \delta\phi^2  
-2\Phi ( \dot\phi \delta\dot\phi + V'\delta\phi ) 
+6\Psi ( \dot\phi \delta\dot\phi - V'\delta\phi ) 
\biggr], 
\end{eqnarray} 
%
where we took the longitudinal gauge and neglected the terms of second 
order in the metric perturbations since they are higher-order in the 
slow-roll parameter. 
Using Eq.(\ref{psi_fo}) and (\ref{phi_fo}), 
the total action becomes 
%
\begin{eqnarray} 
S &=& \int d^5 x \frac{e^{-3\alpha - 3W}}{6\kappa^2} \biggl[ ( 
\Delta\dot\Omega )^2 - \left( \Delta\Omega' \right)^2 
+e^{-2\alpha}\Delta\Omega\Delta^2\Omega  
+\mu^2 e^{2W}\left( \Delta\Omega \right)^2 \biggr] 
\nonumber\\&& 
+\int d^4 x \frac{e^{\alpha}}{6\kappa^2} \biggl[ 
\frac{9}{2}W' \dot {\cal F}^2 
-\frac{3}{2}W' e^{-2\alpha} {\cal F}\Delta {\cal F} 
- W'e^{-4\alpha} \Omega \Delta^2 \Omega \biggr] 
\nonumber\\&& 
+\frac{1}{2} \int d^4 x e^{3\alpha} \biggl[ 
\delta \dot\phi^2 + e^{-2\alpha} \delta\phi \Delta \delta\phi 
-V'' \delta\phi^2  
+2W' {\cal F} e^{-\alpha} (2\dot\phi\delta\dot\phi -V'\delta\phi) 
\nonumber\\&& 
+\dot\phi\delta\dot\phi \, e^{-\alpha} 
\biggl( \ddot\Omega -5H\dot\Omega +4H^2\Omega 
-\frac{5}{3}e^{-2\alpha}\Delta\Omega \biggr) 
+V'\delta\phi \, e^{-\alpha} 
\biggl( \ddot\Omega +H\Omega -2H^2\Omega 
+\frac{1}{3}e^{-2\alpha}\Delta\Omega \biggr) 
\biggr]. 
\label{actionF}
\end{eqnarray} 
%
Taking the variation of this second order action with respect to 
$\Omega$, we get the equation of motion for $\Omega$ in the bulk and the 
junction condition: 
%
\begin{eqnarray} 
e^{-2\alpha} \Delta {\cal F} = \kappa^2 e^{\alpha} 
\dot\phi \delta\dot\phi, 
\label{junc2} 
\end{eqnarray} 
%
where we used the equation of motion for scalar field at the zeroth 
order in the slow-roll parameter. 
From Eq.(\ref{junc2}), we can derive Eq.(\ref{junc_k2b}). 
We also get another junction condition by minimizing the 
action with respect to ${\cal F}$ as 
%
\begin{eqnarray} 
&&3(\ddot {\cal F} +\dot\alpha\dot {\cal F}) +e^{-2\alpha} \Delta {\cal F} = 
\kappa^2 e^{\alpha}( 4\dot\phi \delta\dot\phi -2V' \delta\phi). 
\label{junc1} 
\end{eqnarray} 
%
Combining the junction conditions, we get an evolution equation for ${\cal F}$, 
Eq.(\ref{eqFb}). 

Now we derive the action for $Q$. 
It is useful to notice that the terms contain ${\cal F}$ in 
Eq. (\ref{actionF}) are the same as the 4-dimensional theory if we 
rewrite ${\cal F}$ by $\Psi_4$ using Eq.(\ref{4dpart}). Thus it is convenient to 
define $Q_4$ as 
%
\begin{equation} 
Q_4 = \delta\phi - \frac{\dot{\phi}}{H} 
\Psi_4 =\delta \phi - \frac{\dot\phi}{H} \frac{W'}{2}e^{-\alpha}{\cal F}. 
\label{mukhanov4d} 
\end{equation} 
%
Using Eq.(\ref{junc2}) and (\ref{mukhanov4d}) to express $\dot {\cal F}$, 
$\delta\phi$ and $\delta\dot\phi$ in terms of ${\cal F}$, $Q_4$ and 
$\dot{Q_4}$, we obtain the following action: 
%
\begin{eqnarray} 
S &=& \int d^5 x \frac{e^{-3\alpha - 3W}}{6\kappa^2} \biggl[ ( 
\Delta\dot\Omega )^2 - \left( \Delta\Omega' \right)^2 
+e^{-2\alpha}\Delta\Omega\Delta^2\Omega  
+\mu^2 e^{2W}\left(\Delta\Omega \right)^2 \biggr] 
\nonumber\\&& 
+ \frac{1}{2} \int d^4 x e^{3\alpha} \biggl[ 
\dot{Q_4}^2 + e^{-2\alpha} Q_4 \Delta Q_4 
-(V''+6\dot H) Q_4^2 
\nonumber\\&& 
-\frac{W'}{3\kappa^2} e^{-6\alpha}(\Delta\Omega)^2 
-\dot\phi e^{-\alpha} \biggl(\dddot\Omega -3H^2\dot\Omega +2H^3\Omega 
-e^{-2\alpha}\Delta \left( \frac{5}{3}\dot\Omega -H\Omega \right) 
\biggr) Q_4 
 \biggr]. 
\label{Qb_action} 
\end{eqnarray} 
%
If we express $\cal F$ in Eq.(\ref{mukhanov4d}) by $\Psi$ and $\Omega$ 
using Eq.(\ref{psi_fo}), we immediately get the action in terms of $Q$ 
and $\dot Q$ as 
%
\begin{eqnarray} 
S &=& \int d^5 x \frac{e^{-3\alpha - 3W}}{6\kappa^2} \biggl[ ( 
\Delta\dot\Omega )^2 - \left( \Delta\Omega' \right)^2 
+e^{-2\alpha}\Delta\Omega\Delta^2\Omega  
+\mu^2 e^{2W}\left( \Delta\Omega \right)^2 \biggr] 
\nonumber\\&& 
+ \frac{1}{2} \int d^4 x e^{3\alpha} \biggl[ 
\dot{Q}^2 + e^{-2\alpha} Q \Delta Q 
-(V''+6\dot H) Q^2 
\nonumber\\&& 
-\frac{W' e^{-6\alpha}}{3\kappa^2}(\Delta\Omega)^2 
+\frac{\dot\phi e^{-3\alpha}}{3H} Q \Delta \biggl( 
\ddot\Omega -H\dot\Omega -e^{-2\alpha}\Delta\Omega \biggr) 
 \biggr]. 
\label{Q_action} 
\end{eqnarray} 
%
We shoud note that the slow-roll approximation was 
used to derive Eq.(\ref{Qb_action}) and (\ref{Q_action}). 
The equation of motion for $Q$ can be easily derived from this action: 
%
\begin{eqnarray} 
&&\ddot Q +3H\dot Q -e^{-2\alpha}\Delta Q = -(V'' +6\dot H)Q 
+\frac{\dot\phi e^{-3\alpha}}{6H}\Delta \biggl( 
\ddot\Omega -H\dot\Omega -e^{-2\alpha}\Delta\Omega \biggr). 
\end{eqnarray} 
%
Of course, this agrees with Eq.(\ref{Qevolution}). 
We can also derive the junction condition for $\Omega$:
\begin{equation}
e^{-2 \alpha} \Delta {\cal F} = \kappa^2 e^{\alpha}
\dot{\phi} \dot{Q}.
\label{couple}
\end{equation}

The action (\ref{Q_action}) is the main result of this 
paper. This action describes the coupling between 
the matter fields on the brane and gravitational 
fields in the bulk. Although the 5-dimensional gravitational
fields are very complicated, the final action is very simple
if we use the master variable. Essentially the system is 
described by two scalar fields. One is living on the brane 
and the other 5-dimensional scalar field is living in the 
bulk and they are coupled with each other on the brane. 

In general it is very difficult to quantize this coupled 
system of the bulk and the brane. However, from the 
solution for $\Omega$ (\ref{solution_omega}), we can 
see that the coupling term is suppressed by the 
slow-roll parameter. Thus it is possible to solve
the equations perturbatively. At the 0-th order, 
the scalar field perturbation $Q_0=\delta \phi_0$ decouples.
Then by quantizing $Q_0$, we can determine the coefficient $C_1$ and $C_2$ in 
Eq.(\ref{F_solution}) as 
%
\begin{eqnarray} 
C_1 = \kappa^2 \frac{i\dot\phi}{\sqrt{2k} H}, C_2 = i C_1, 
\end{eqnarray} 
%
where we chose a standard Bunch-Davis vacuum.  
The master variable $\Omega$ couples to $Q_0$ via the 
junction condition Eq.(\ref{couple}). Then the normalization
of $\Omega$ is also determined from the solutions Eq.(\ref{solution_omega}). 
There are also contributions from the normalizable KK modes 
with $m>3H/2$ which satisfy the vacuum junction condition ${\cal F} = 
0$. The contribution of the KK continuum with $m>3H/2$ to the bulk 
perturbations are the same as the case of the vacuum brane, 
so the normalization is determined as in Sec.\ref{quantum_heavy}. 
With these KK modes as well as the discrete modes induced by the 
scalar field on the brane, we can calculate the correction term $J$, which is 
needed to know the behavior of $Q_1$. 

Because there is a contribution from the infinite ladder of 
modes in $J$, the contribution of $J$ could be significant 
especially at high energies $H/\mu \gg 1$ on small scales. 
In this case we should carefully reexamine our perturbation scheme. 
A quantification of the 5-dimensional corrections is beyond the scope of 
our present paper and we hope to report results in future publications.  

\section{SUMMARY AND DISCUSSION} 
\label{summary}

In this paper, we considered the quantum scalar perturbation about a de 
Sitter brane in a 5-dimensional AdS bulk spacetime. 
We first introduced the gauge-invariant master variable $\Omega$ for 
gravitational perturbations in the AdS spacetime found by 
Mukohyama \cite{Mukohyama}. 
We then derived its second order action which is needed to 
perform the quantization of the master variable. 

In the case of a vacuum single de Sitter brane, there is a 
continuum of normalizable KK modes with $m>3H/2$. 
These bulk perturbations are felt on the brane through the projection of 
the perturbed 5-dimensional Weyl tensor $\delta E_{\mu\nu}$. 
We calculated the vacuum expectation value of its effective energy 
density by using the second order action for $\Omega$. 
A light radion mode with $m=\sqrt{2}H$ also satisfies the junction 
conditions on the branes and is normalizable for a two branes model. 
The energy density of $\delta E_{\mu\nu}$ due to this radion mode is 
also calculated and is shown to agree with the previous 
results obtained by Gen and Sasaki \cite{GenSasaki2}. 

Next we considered the case where there is a scalar field perturbation 
on a single de Sitter brane. 
The bulk perturbations can be solved using the slow-roll approximation 
\cite{koyama04}. 
As shown in \cite{koyama04}, 
the $m^2 = 2H^2$ mode together with the zero-mode and an infinite ladder 
of discrete tachyonic modes become normalizable. 
There are also contributions from the continuum of the KK modes which 
satisfy the vacuum boundary condition on the brane ${\cal F} = 0$. 
These bulk perturbations introduce corrections to the scalar 
type perturbations of the standard 4-dimensional cosmology, 
as is seen from Eq.(\ref{Qevolution}). 
To calculate the amplitude of scalar perturbations on the brane, 
we need to compute the vacuum expectation value of Mukhanov-Sasaki 
variable $Q$. Then we derived the sencond order action
(\ref{Q_action}) for $Q$ coupled to 5-dimensional bulk perturbations. 
Because the coupling term between $Q$ and the master variable $\Omega$
is suppressed by a slow-roll parameter, we can solve the system 
by a slow-roll expansion. 

The action (\ref{Q_action}) describes the essential features of 
scalar perturbations in the brane world. The scalar field 
perturbation on the brane inevitably produces the bulk gravitational 
perturbations, which back-react to the perturbations on the brane. 
A detailed analysis of this coupled bulk-brane system is very important 
in order to find brane wolrd signatures from inflation and we 
hope to come back to this issue in future publications. 

\section*{Acknowledgments} 

H.Y. acknowledges valuable discussions with Shinji Mukohyama. 
The works of H.Y. and K.K. are supported by Grant-in-Aid for JSPS 
fellows. 


\end{document}